\documentclass[aps,pre,twocolumn,groupedaddress,floatfix]{revtex4-1}

\usepackage{graphicx}
\usepackage{color}
\usepackage{amssymb,amsfonts,amsmath}
\usepackage[english]{babel}
\usepackage{enumerate}

\begin{document}

\title{Driven diffusion against electrostatic or effective energy barrier 
across $\alpha$-Hemolysin}

\author{Patrizio Ansalone}
\thanks{There authors contributed equally to this work}
\affiliation{Istituto Nazionale di Ricerca Metrologica, Strada delle Cacce 91, Torino, IT-10135, Italy}
\author{Mauro Chinappi}
\thanks{There authors contributed equally to this work}
\affiliation{Center for Life Nano Science@Sapienza, Istituto Italiano di
Tecnologia, Via Regina Elena 291, 00161, Roma, Italia}
\author{Lamberto Rondoni}
\affiliation{Scienze Matematiche, Politecnico di Torino Corso Duca degli 
Abruzzi 24, 
Torino, IT-10129 Italy; 
INFN, Sez. di Torino, Via P. Giuria 1, Torino IT-10125, Italy}
\author{Fabio Cecconi}
\email{fabio.cecconi@roma1.infn.it}
\affiliation{CNR-Istituto dei Sistemi Complessi UoS ``Sapienza'',
Via dei Taurini 19, 00185 Roma (Italy)}

\date{\today}

\begin{abstract}
We analyze the translocation of a charged particle across an
$\alpha$-Hemolysin ($\alpha$HL) pore in the framework of a driven diffusion over
an extended energy barrier generated by the electrical charges of the
$\alpha$HL.
A one-dimensional electrostatic potential is extracted from the full 3D solution of 
the \emph{Poisson}'s equation.
We characterize the particle transport under the action of a constant forcing 
by studying the statistics of the translocation time.
We derive an analytical expression of translocation time
average that compares well with the results from Brownian dynamic simulations
of driven particles over the electrostatic potential.
Moreover, we show that the translocation time distributions
can be perfectly described by a simple theory
which replaces the true barrier by an equivalent structureless square barrier.
Remarkably our approach maintains its accuracy also for low-applied voltage
regimes where the usual inverse-Gaussian approximation fails.
Finally we discuss how the comparison between the
simulated time distributions and their theoretical prediction
results to be greatly simplified when using the notion of the empirical
Laplace transform technique. 
\end{abstract}

\keywords{driven diffusion, first-passage process, $\alpha$-Hemolysin, empirical Laplace transform.}

\maketitle

\section{Introduction}
\label{sec:intr}
Theoretical and computational approaches to driven or spontaneous migration of molecules
through a biological pore are often based on idealized models whereby pores are represented
as passive channels imposing a spatial confinement \cite{matysiak2006,muthu2008,huang2008translocation,
Cecconi_2011,bacci2012role} that basically results into the presence of an entropic barrier
\cite{berezhkovskii02,muthuBook}.
While this raw geometrical picture may be appropriate to investigate a general principles
of the transport of neutral species across solid state nano-channels,
it turns out to be drastic for biological pores whose chemical composition
is known to affect the translocation mechanism of charged molecules. Both experiments and simulations have,
indeed, revealed that pore charges strongly
influence the translocation under the effect of driving forces
\cite{movileanu05inter,wong2010polymer,muthukumar2006simulation,
barkema2013,trick2014designing,mereuta2014slowing}
Therefore, the electrostatic interaction between a pore and the charged molecules
cannot be neglected in reasonably realistic phenomenological models of
translocation.

From a theoretical perspective, the passage
of molecules through a nanopore can be viewed as the overcoming a free-energy
barrier determined by the physical properties of the pore and molecule system.
This problem is commonly tackled as a driven-diffusion process in the presence of a given
free-energy landscape
\cite{lubensky1999,ammenti2009statistical,pelizzola2013nonequilibrium,im2002ion,bacci2013protein} ,  which
under suitable approximations, leads to solving a one-dimensional driven diffusion \emph{Smoluchowski}
equation \cite{gardiner1985handbook}, with appropriate initial and boundary conditions.

The main purpose of this paper is to analyze the driven diffusion of a charged particle in the
presence of realistic electrostatic potential.
As a relevant example we selected the $\alpha$HL pore,
a biological pore
widely employed in nanopore technology
\cite{song1996structure,madampage2012nanopore,oukhaled2012transport,oukhaled2007unfolding,
nivala2013unfoldase,rodriguez2013multistep,asandei2015placement,di2015all,asandei2015acidity}
that spontaneously self assembles into an heptameric
channel that inserts itself  into a lipid bilayer.
First, we compute numerically the 3D electrostatic potential generated by the $\alpha$HL then
we interpolate the effective one-dimensional profile along the pore channel.
In the Smoluchowski driven-diffusion picture such a 1D-potential corresponds to an energy barrier
that a charged particle has to overcome in order to cross the pore, with the simplified assumption that the
translocating particle does not perturb the charge distribution and does not affect the electrostatics of the $\alpha$HL.
In this context, the translocation of a charged monomer is assimilated to a first passage process of a
Brownian charged particle entering the pore in one side and reaching the opposite one.
The statistical properties of the translocation are then obtained either by direct
numerical integration of the Langevin equation or via the, computational
or analytical solution of the
Smoluchowski equation \cite{gardiner1985handbook}.

In this study we derive an analytical expression for the average
passage-time as a function of the external load
for generic 1D barrier.
In addition, for the specific case of a square barrier we find
a closed expression for the Laplace transform of the passage
time distribution.
Interestingly, the distributions numerically obtained, with the 1D potential
are well matched by the theoretical one with an equivalent square barrier.
As a final remark, our results suggest that the fitting procedure 
of theoretical first passage-time distributions to numerical 
or experimental data can be conveniently carried out  
via the empirical Laplace technique~\cite{henze2002goodness}.

The present paper is organized as follows. 
Section \ref{sec:ele} describes the electrostatic model adopted in a ideal free salt environment.
In section \ref{sec:smo}, we briefly introduce the \emph{Smoluchowski} description for driven diffusion,
we derive the analytical solution for the average passage time for a generic free-energy profile, and we obtain
the Laplace transform of the translocation time distribution for the square barrier. Numerical results for
the first passage time distributions are reported and discussed in section~\ref{sec:Results}.

\section{Electrostatic Potential Barrier} \label{sec:ele}

\begin{figure}[tbp]
\includegraphics[clip=true,keepaspectratio,width=\columnwidth]
{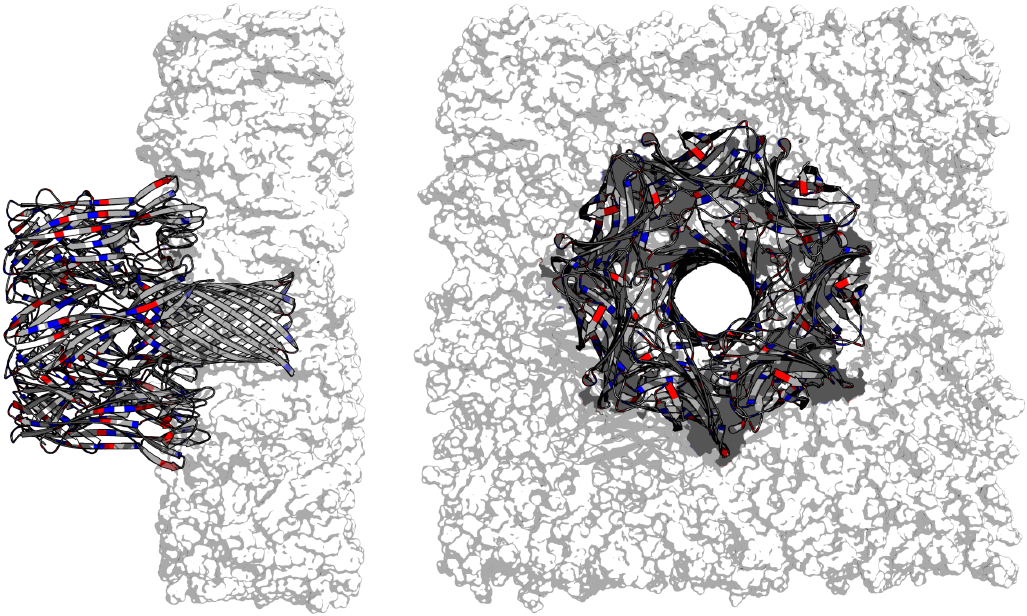}
\caption{The mushroom-shaped protein $\alpha$HL consists
of a vestibule with a diameter $\approx 46\mathring{A} $ at its widest 
section and $50\mathring{A} $ in length. The vestibule is linked to a 
$\beta$-barrel transmembrane whose diameter is $\approx 25\mathring{A} $ and
$50\mathring{A} $ in length. The interface between the vestibule
and the $\beta$-barrel domains shows a bottleneck;
in this region the pore is $\approx14\mathring{A} $ wide at its narrowest point.}
\label{fig:alpha}
\end{figure}

\begin{figure}[tbp]
\includegraphics[clip=true,keepaspectratio,width=\columnwidth]
{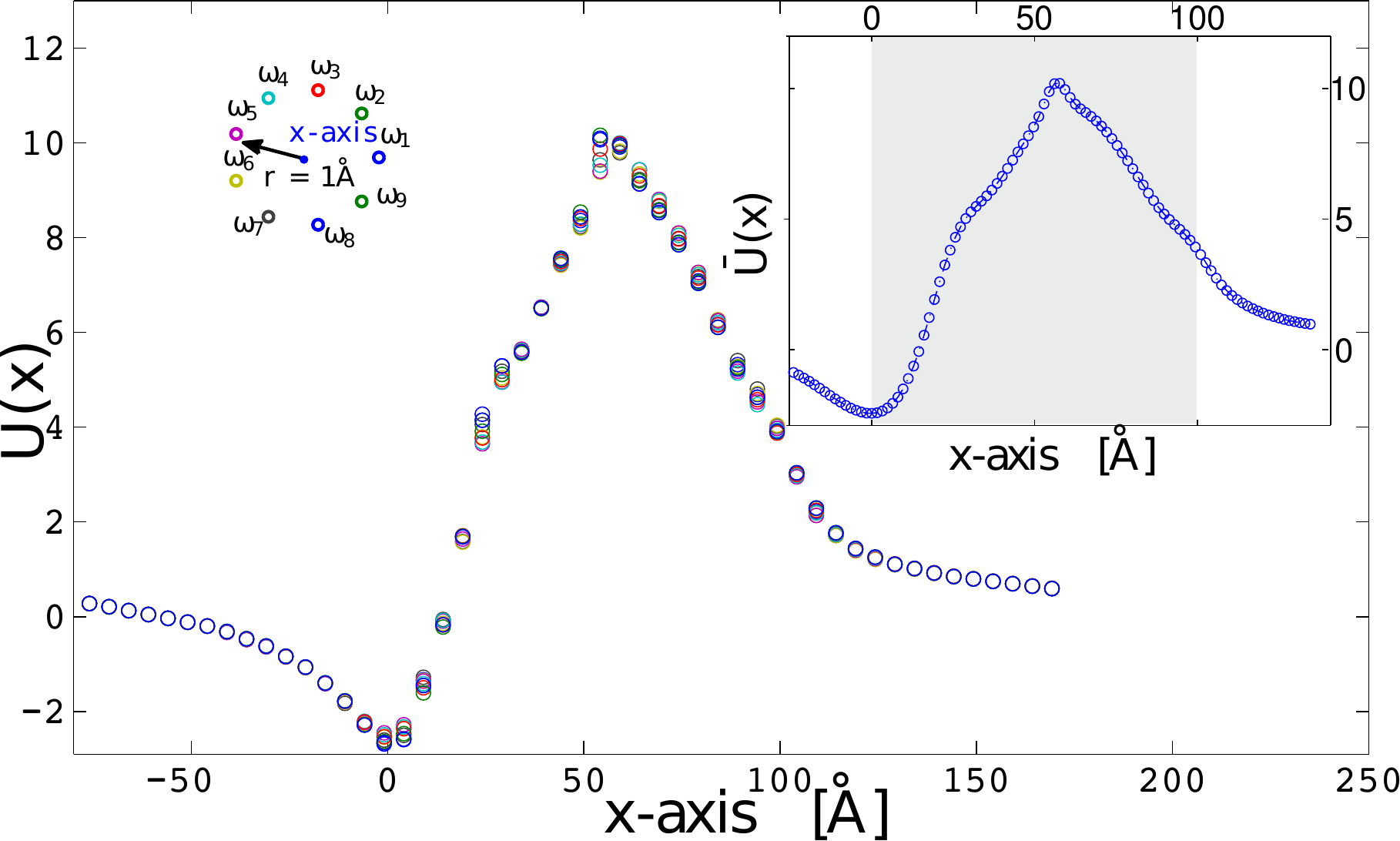}
\caption{Dimensionless unit $k_{B}Te^{-1}$ of the electrostatic potential 
along the $x$-axis of the
$\alpha$HL. The behavior of the electrostatic potential is plotted
taking into account nine different lines of sight, each one at a $1\mathring{A}$
distance from the $x$-axis. The highest value in the electrostatic potential
is located approximately $40\mathring{A} $ from the origin of the
reference frame (i.e.\ the cap of the $\alpha$HL).
It corresponds to the location of a group of Lys$^{147}$
residues. The average of the electrostatic potential is shown in the inset.
The grey shaded area corresponds to the pore region.
\label{Fig:Electrostatic_Potential_Hemolysin}
}
\end{figure}

This section presents the computation of the electrostatic potential inside and outside the
($\alpha$HL) pore  in a free salt medium.
The $\alpha$HL length is approximatively $L = 100\mathring{A} $,
and it is subdivided in several regions, figure \ref{fig:alpha}. The $\beta$-barrel region
(it is embedded in the lipid membrane) has a diameter of  $\simeq 20\mathring{A}$ (based on backbone),
the vestibule is $~46\mathring{A}$ in diameter, and the narrowest section, located between the
vestibule and the $\beta$-barrel, is approximatively $14\mathring{A}$ wide. The pore is
embedded in a lipid membrane approximatively of 40$\mathring{A}$
thickness. In this study $\alpha$HL is aligned
along the $x$-axis. The \emph{Poisson} and  \emph{Laplace}
equations are numerically solved in order to calculate the electrostatic potential $V(\bf{r})$.
\begin{equation} \left\{
\begin{array}{ll}
\nabla^2 V=0&\rm{in}\; \Omega_{\rm{H_2O}}, \Omega_{\rm{m}}\subset\mathbb{R}^3 \\
\displaystyle \nabla^2 V=-\sum\limits_{i=1}^{N_q}\frac{q_i}{\varepsilon_p\varepsilon_0}
\delta(\bf{r}-\bf{r}_i)&\rm{in}\;\Omega_{\rm{p}}\subset\mathbb{R}^3 \; .
\end{array}
\right.
\label{Eqn:Poisson}
\end{equation}
Here, $\Omega_{\rm{H_2O}}$ is the solvent region occupied by the water, $\Omega_{\rm{m}}$ is the lipid
membrane region and $\Omega_{\rm{p}}$ is the region occupied by $\alpha$HL,
while $\varepsilon_0$ and $\varepsilon_p$ are the vacuum and pore  dielectric constants, respectively.
The $N_q$ point charges $q_i$ in Eq.~\eqref{Eqn:Poisson} are obtained from the PDB structure
of $\alpha$HL (pdb code: 7AHL) \cite{song1996structure}, the protonation state for
residues are determined
at pH 7.0 employing the PDB2PQR pipeline~\cite{dolinsky2007pdb2pqr} and the AMBER99 force
field~\cite{wang2000well}.
Briefly in order to simulate the entire system we carried out a dummy
calculations with APBS-FETK software libraries  \cite{holst2001adaptive} to create 3D
dielectric environment and the geometry
of a previously equilibrated lipid membrane.
The dielectric constant of each region of the structure is homogeneous, the values
are chosen as follows: in $\Omega_{\rm{H_2O}}$ the standard relative dielectric is $\varepsilon_{\rm{H_2O}}=78.54$.
For the pore $\varepsilon_{p}= 4$ and  $\varepsilon_{m}= 2$  for the lipid membrane region.
The system of equations is  solved using the APBS-FETK 
\cite{holst2001adaptive,ansalone2010,ansalone2011}.
The simulation is run assuming a bounding box equal to
$321\mathring{A} \times321\mathring{A} \times321\mathring{A}$
with  a fine grid of 1 grid nodes/$\mathring{A}$.
The eq. \eqref{Eqn:Poisson} is numerically solved enforcing Dirichlet boundary condition and continuity
of the potential and normal electric displacement across the interfaces among  solvent, lipid membrane and $\alpha$HL regions.
To obtain a 1D profile of the electrostatic potential
as a function of the $x$-coordinate from the 3D electrostatic potential, we use the following procedure. Firstly we chose
nine directions of sight, labelled by $\omega_i$ in figure \ref{Fig:Electrostatic_Potential_Hemolysin}, at a distance
of $1\mathring{A} $ from the $x$-axis of the pore, then we interpolate the
electrostatic potential for each one of them. Since the potential appears to be only slightly dependent on the position
$\omega_i$, inside the pore, we take the average as electrostatic barrier experienced by a charged
particle translocating through the pore (inset of figure \ref{Fig:Electrostatic_Potential_Hemolysin}).
Finally the average electrostatic potential
$\bar U(x)$
is fitted with a multi-Gaussian function.
The figure \ref{Fig:Electrostatic_Potential_Hemolysin}
shows the behavior of the electrostatic potentials along the $x$-axis of the
$\alpha$HL pore. 
The peak in the electrostatic potential is mostly
localized between the seven Lys$^{147}$, these residues form a ring
around the narrowest part of the $\beta$-barrel region. In contrast, the electrostatic
effects arising from Asp$^{127}$ and Asp$^{128}$ (negatively charged) are partially
compensated by Lys$^{131}$ (positively charged), these residues flatten the peak of
the electrostatic potential at the end of the $\beta$-barrel region
($x \approx 100 \mathring{A}$).
These results are in agreement with previous studies.\cite{simakov2010soft}

\section{Smoluchowski driven diffusion \label{sec:smo}}
\begin{figure}[tbp]
\includegraphics[clip=true,keepaspectratio,width=\columnwidth]
{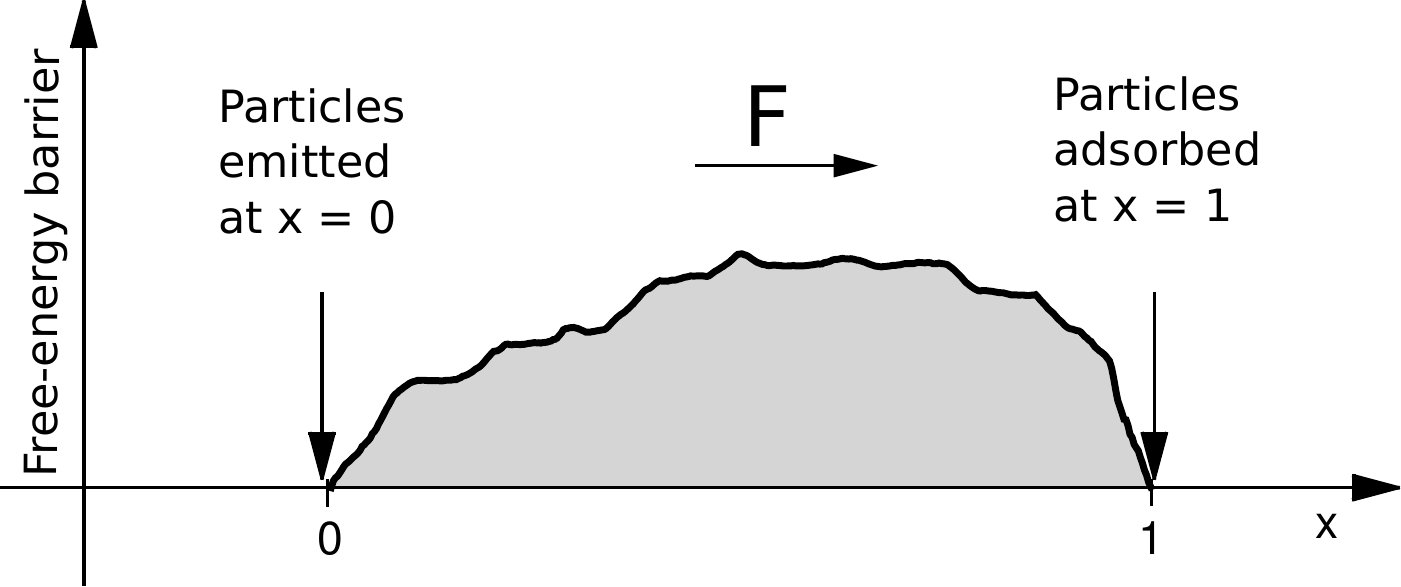}
\caption{Sketch of the 1-D driven-diffusion model.
Particles are emitted at the pore entrance
$x=0$ and are adsorbed at the pore exit ($x=1$). Translocation requires
the overcoming of the free-energy barrier with the help of an external
force $F$ acting along the $x$ direction. \label{fig:barr}}
\end{figure}

Here, we describe the one-dimensional \emph{Smoluchowski} equation for
a positively charged particle $q$ driven by a constant electrical
field $E_x$ along the $x$-axis, $F = q E_x$ in the presence of a generic
energy barrier.
The probability $P(x,t)$ for a particle to be in the position $x$ at time $t$
satisfies the conservation equation
\begin{equation}
\label{eqn:conservative}
\frac{\partial P}{\partial t}
=-\frac{\partial J}{\partial x}\,,
\end{equation}
where $J(x,t)$ is the flux of
probability density.
In the present case, Eq.(\ref{eqn:conservative}) takes the form
\begin{equation}
\label{eqn:constitutive}
J(x,t) = -D \frac{\partial
P}{\partial x}-\mu P\frac{\partial
\bar U(x)
}{\partial x} + \mu PF\;,
\end{equation}
where $\mu$ and $D = \mu k_B T$ are the particle mobility and diffusion coefficient respectively,
with $T$ the temperature and $k_B$ the Boltzmann constant. The function
$\bar U(x)$
denotes the barrier
profile due to the pore as defined in the previous section
and $F$ is the applied bias electric force
\textcolor{black}{acting over the whole domain $[-\infty,L]$}. 
The pore occupies the
region $x \in (0,L)$. Particles are emitted at the pore entrance, $x=0$, at time $t = 0$,
and are later adsorbed at $x = L$, which implies the boundary condition
$P(L,t) = 0$, see Fig.~\ref{fig:barr}.
In the following, unless differently specified, we use the natural
dimensionless variables:
$\tilde x=L^{-1}x, \tilde t=DL^{-2}t$, $\tilde F = L F/ k_B T$.
In these coordinates, the pore region $[0,L]$ is rescaled to the interval
$[0,1]$. For sake of notation simplicity, we omit the tilde, so that the
dimensionless Smoluchowski equation reads:
\begin{equation}
\label{eqn:Smoluchowski}
\frac{\partial
P}{\partial{t}}= \frac{\partial }{\partial x}
\left[ \frac{\partial P}{\partial x} +
P \frac{\partial
\bar U(x)
}{\partial x} - F P
\right] \,.
\end{equation}
Integrating the solution $P(x,t)$ of Eq.~\eqref{eqn:Smoluchowski},
we obtain the probability $S(t)$ that at time $t$ the particle has
not yet translocated
\begin{equation}
S(t)=\int_{-\infty}^{1} P(x,t ){d}x\;.
\label{eq:Survival}
\end{equation}
This is also called the survival probability, as in our model, it is the probability that the particle has
not been absorbed at $x=1$. Accordingly, the probability to be absorbed, i.e.\ to exit the pore after the time
$t$, is $P_{\mathrm{out}}(t) = 1-S(t)$. Therefore the distribution of first passage times $\psi(t)$ is
simply obtained as
\begin{equation}
\label{eqn:dotprob}
\psi(t)=\frac{d P_{\mathrm{out}}(t)}{dt}= -\frac{{d} S(t)}{dt}.
\end{equation}
Using the Smoluchowski Eq.~\eqref{eqn:Smoluchowski}, we can express 
$\psi(t)$ in terms of the probability current
at the absorbing boundary
\begin{equation}
\psi(t)=J(1,t)=-\frac{\partial P}{\partial x}\bigg|_1-P(1,t)\,
\frac{\partial
\bar U(x)
}{\partial x}\bigg|_1+FP(1,t)\,,
\label{eq:psi1}
\end{equation}
thus the probability current, through the solution of Eq.~\eqref{eqn:Smoluchowski},
determines $\psi(t)$.
The knowledge of $\psi(t)$ allows us to derive all moments of the distribution and, in particular,
the average translocation time:
\begin{equation}
\label{eqn:meantime}
\tau=\int_{0}^{\infty} dt
\;t\;\psi(t)=\int_{-\infty}^{1}dx \int_{0}^{\infty}dt P(x,t)
\end{equation}
where the second equality follows from 
Eqs.~(\ref{eqn:dotprob},\ref{eq:Survival}) and from an
integration by parts. Appendix \ref{app:ana} shows that,  
for a generic barrier of shape $\bar U(x)$,
the Smoluchowski equation in the domain
$[-\infty,1]$ with initial condition $P(x,0) = \delta(x)$ provides the 
expression:
\begin{equation}
\tau = \frac{M_{+}(F)}{F} + M_0(F)~,
\label{eq:tau}
\end{equation}
where the functions $M_{+}(F)$ and $M_{0}(F)$ are given by
\begin{equation}
\begin{array}{l}
\displaystyle
M_{+}(F)  = \int_0^1 dx\;e^{G(x)} \\
\\
\displaystyle
M_{0}(F)  = \int_0^1 dx\;e^{-G(x)}\int_{x}^1 dy\; e^{G(y)}\;,
\end{array}
\label{eq:gliM}
\end{equation}
with
$G(x) = \bar U(x) - F x$.
Interestingly, in the limit $F \to 0$, the second term of eq. \eqref{eq:tau}
becomes negligible and
\begin{equation}
\tau \simeq \frac{M_{+}(0)}{F}  \ ,
\label{eq:tau0}
\end{equation}
as we will see in the following, this allows an equivalent square
barrier to be defined with height
\begin{equation}
\phi = \log M_{+}(0)  \, .
\label{eq:phiequiv}
\end{equation}

While the explicit value of $\tau$ can be obtained numerically,
at least, computing the relevant integrals,
the explicit expression of $\psi(t)$ for an arbitrary barrier cannot be given. Indeed, Eq.~\eqref{eq:psi1}
requires the full solution of the \eqref{eqn:Smoluchowski} which,
in general, cannot be worked out analytically.
Consequently, one must either resort to direct numerical simulations or
introduce simplifying approximations.

A simple but meaningful approximation, amounts to replacing the actual potential barrier with a square
profile which allows us to derive a closed analytical form for the Laplace transform $\hat\psi(s)$
of $\psi(t)$, and which appears to be ``equivalent'' to the real barrier for calculations of our concern.
Indeed, as detailed in \ref{app:ana}, the Laplace transform of the Smoluchowski equation yields the result:
\begin{widetext}
\begin{equation}
\hat\psi_{\phi}(s)=\frac{2 e^{F/2} A(F,s)} {\left[2e^{\phi} A(F,s)+
F(1-e^{\phi}) \right] \sinh[A(F,s)] + 2A(F,s)\cosh[A(F,s)]}
\label{eq:phis}
\end{equation}
\end{widetext}
where $2A(F,s) = \sqrt{F^2 + 4s}$, and $\phi$ is the barrier height, i.e.\
$\bar U(x)=\phi$
for $x \in [0,1]$ and
$\bar U(x)=0$, $x \in [-\infty,0]$. As we shall see in Section~\ref{sec:Results}, the expression of $\hat \psi(s)$
is very useful in the analysis of the Empirical Laplace transform ~\cite{henze2002goodness} of the
translocation times obtained from direct simulations, even without  
explicit inversion.
Eq.~\eqref{eq:phis} reduces to the well known \emph{Inverse Gaussian} (IG) distribution obtained in
the vanishing barrier limit,
$\phi = 0$:
\begin{equation}
\hat{\psi}_{IG}(s) = \exp\bigg\{\frac{1}{2}
\bigg(F - \sqrt{F^2 + 4s}\bigg)
\bigg\}~,
\label{eq:IGlaplace}
\end{equation}
which can be easily inverted to give
\begin{equation}
\psi_{IG}(t) = \frac{1}{\sqrt{4\pi t^3}}\,
\exp\bigg\{-\frac{(1- F t)^2}{4 t}\bigg\}~,
\label{eq:invGauss}
\end{equation}
customarily considered a useful guide to interpret data of voltage-driven
translocation experiments in high-voltage regimes~\cite{ling2013distribution}.
\textcolor{black}{At the end of Appendix \ref{app:LapInv}, we show that, 
though the inversion of Eq.\eqref{eq:phis} leads to a quite involved expression, 
the large-time behaviour of the arrival time PdF can be easily worked out
via a saddle-point method (see Eq.\eqref{eq:decay}) and reads 
\begin{equation}
\psi_{\phi}(t) 
\sim \exp\bigg\{-\frac{F^2 e^{\phi}}{(1+e^{\phi})^2}\;t\bigg\}\;.
\label{eq:expcoeff}
\end{equation}
Interestingly, the exponential decay rate is controlled by the height 
of the equivalent barrier and, when $\phi\to 0$, it is consistent with 
the Inverse Gaussian behaviour $\exp(-F^2t/4)$.}

Using the relation between the derivatives of the Laplace transform calculated at $s=0$ with the momenta of the function,
$(-1)^{n} \hat \psi^{(n)}(0) = \int_0^{\infty} dt\;t^n \psi (t)$,
where $(n)$ indicates the n-th derivative, we can obtain all the
momenta of $\psi$, in particular, the average residence time reads
\begin{equation}
\tau = \frac{F+\left(e^\phi - 1\right)
\left(1-e^{-F}\right)} {F^2} \,.
\label{eq:tau3}
\end{equation}
Notice that Eq.\eqref{eq:tau3} can be also directly derived
by Eq.\eqref{eq:tau}.
Interestingly, in the limit of $F \to 0$ \eqref{eq:tau3} reduces to
\begin{equation}
\tau = \frac{e^\phi}{F} \, ,
\label{eq:tauS0}
\end{equation}
i.e. again a $1/F$ behavior as in the $F \to \infty$ limit.
By comparison of eq.  \eqref{eq:tauS0} and eq. \eqref{eq:tau0} it results that,
in the limit $F \to 0$, any barrier can be described
by an equivalent square barrier the correspondence being set
via the following relation
\begin{equation}
\exp(\phi) = M_{+}(0)  = \int_0^1 dx\;e^{U(x)} \, .
\label{eq:phi}
\end{equation}

\section{Numerical results \label{sec:Results}}

The electrostatic barrier derived in Section \ref{sec:ele} is used to study
numerically the translocation of a positive unit charge across the $\alpha$HL
under the action of a constant electric field $E$.
We generate a continuous version of 1D average electrostatic potential by a
multi Gaussian fit.
\begin{equation}
\label{eq:fit}
\bar U(x)  =
\sum_{k=0}^{7} \bar U_k \exp\bigg[-\bigg(\frac{x-x_k}{c_k}\bigg)^2\bigg]\;.
\end{equation}
The set of coefficients is reported in table \ref{tab:U}.
\begin{table}[]
\caption{multi-Gaussian fit coefficients, eq. \eqref{eq:fit},
with a $95\%$ confidence intervals.
$\bar U_k$ and in $k_B T e^{-1}$ while $x_k$ and
$c_k$ are in \AA.}
\begin{center}
\begin{tabular}{ | c | c | c | c |}
  \hline
  $k$  & $\bar U_k$ & $x_k$ & $c_k$	\\ \hline
  \hline  
  0  &-0.547 & -50.000 &  20.260 \\ \hline
  1  &-4.000 &    1.000 &  29.230 \\ \hline
  2  & 2.510 &  18.940 &   11.110 \\ \hline
  3  & 0.729 &  49.243 &    0.003  \\ \hline
  4  & 1.047 &  50.432 &     4.396 \\ \hline
  5  & 8.340 &  52.766 &   35.590 \\ \hline
  6  & 1.112 &  68.610 & 125.000 \\ \hline
  7  & 0.598 &  92.130 &     7.732 \\ \hline
  \end{tabular}
  \label{tab:U}
  \end{center}
\end{table}
The dynamics of the charged particle is
described by the overdamped Langevin equation:
\begin{equation}
\dot{x} = F - \frac{\partial
\bar U(x)
}{\partial x}+\sqrt{2}\eta(t)
\label{eq:lang1}
\end{equation}
with $\eta(t)$ a Gaussian noise, with $\langle \eta(t)\rangle = 0$ and
$\langle \eta(t)\eta(t')\rangle = \delta(t-t')$. 
\textcolor{black}{After ensemble averaging, 
this Langevin approach is equivalent to the Smoluchowski formulation
\eqref{eqn:Smoluchowski}.\cite{gardiner1985handbook} 
In the Langevin formulation \eqref{eq:lang1}, we assume the 
friction exerted by the solvent large enough to overwhelm the inertial 
terms (overdamped regime).}
We integrated Equation~\eqref{eq:lang1} numerically
via a second order stochastic Runge-Kutta algorithm 
\cite{honeycutt1992stochastic},
for an ensemble of $M = 10^5$ independent particles emitted at the pore
entrance ($x = 0$) at $t = 0$ and adsorbed at $x = 1$.
\textcolor{black}{We notice that particles are allowed to explore 
the whole domain $[-\infty,1]$}.
As it is customary the translocation time in our simulations 
is the time of first arrival at the absorbing boundary 
$t_{\mathrm{Tr}} = \min\{t\in [0,T_w] \,:\, x(t) = 1 \}$.

\begin{figure}[tbp]
\includegraphics[clip=true,keepaspectratio,width=\columnwidth]
{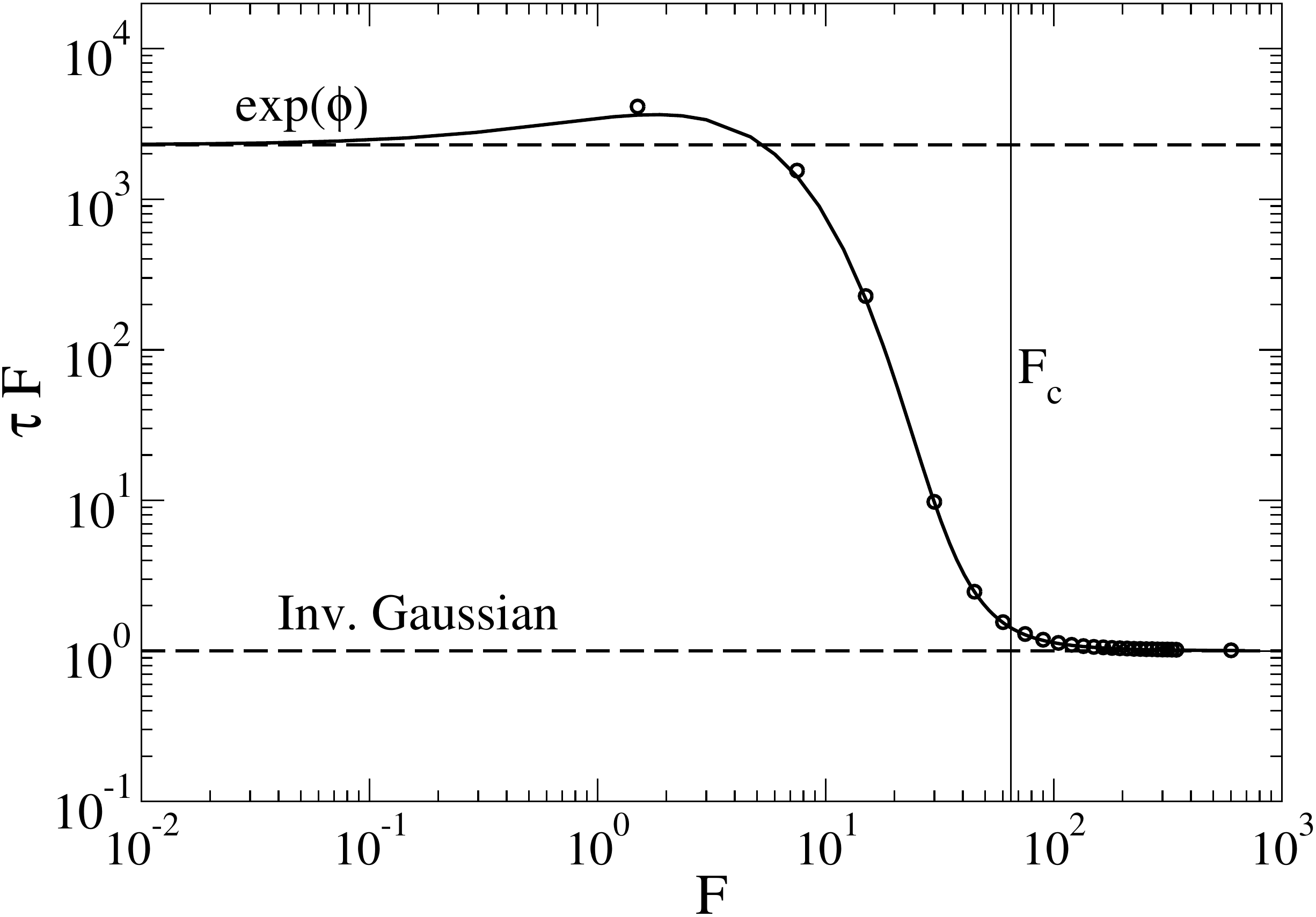}
\caption{Behavior of the average translocation time $\tau$ versus $F$.
The solid line is the analytic result Eq.~(\ref{eq:tau}) while the points
refer to numerical simulations.
For high forcing ($F > F_c$) the curve collapses on the
ballistic (inverse Gaussian) prediction $\tau = 1/F$.
The higher dashed line indicates the $F=0$ limit, Eq.\eqref{eq:tauS0}.
The vertical line marks the stability threshold $F_c = \max_{x}{U'(x)}$
at which the profile $G(x) = \bar U(x) - F_c x$
looses its stable minimum given by the solution $U'(x)=F$.
\label{fig:time}}
\end{figure}

Figure~\ref{fig:time} illustrates the average translocation time $\tau$
over the $M$ trajectories as a function of the external load $F$ (cf.\ Section~\ref{sec:smo}).
The points are the results of the Brownian dynamic simulations and  
for comparison, the analytical curve Eq.~\eqref{eq:tau} is also plotted,
demonstrating the agreement with the numerical data.
Different regimes are apparent in the behavior of $\tau(F)$.
For high forcing, the typical Inverse Gaussian behavior,
$\tau = F^{-1}$, is recovered (lower horizontal dashed line).
The stability threshold $F_c$ for the onset of this
ballistic-like regime can be roughly estimated as
$F_c = \max_{x}\bar U'(x) \simeq 64.5$,
that is the minimal $F$-value after which the barrier of the 
tilted profile $G(x) = \bar U(x) -F x$ disappears 
\cite{DudkoPRL06,ReimannJPCM15,plata}. 
Below $F_c$, $\tau$ abruptly increases as the barrier crossing turns to be
mainly thermally activated.
The vertical line in Fig.~\ref{fig:time} represents the threshold
$F_c$ separating a thermally activated from a ballistic-like regime.
Moreover, it is apparent that for low forcing the
exit-time goes as Eq.\eqref{eq:tauS0}.

\begin{figure}[tbp]
\includegraphics[clip=true,keepaspectratio,width=\columnwidth]
{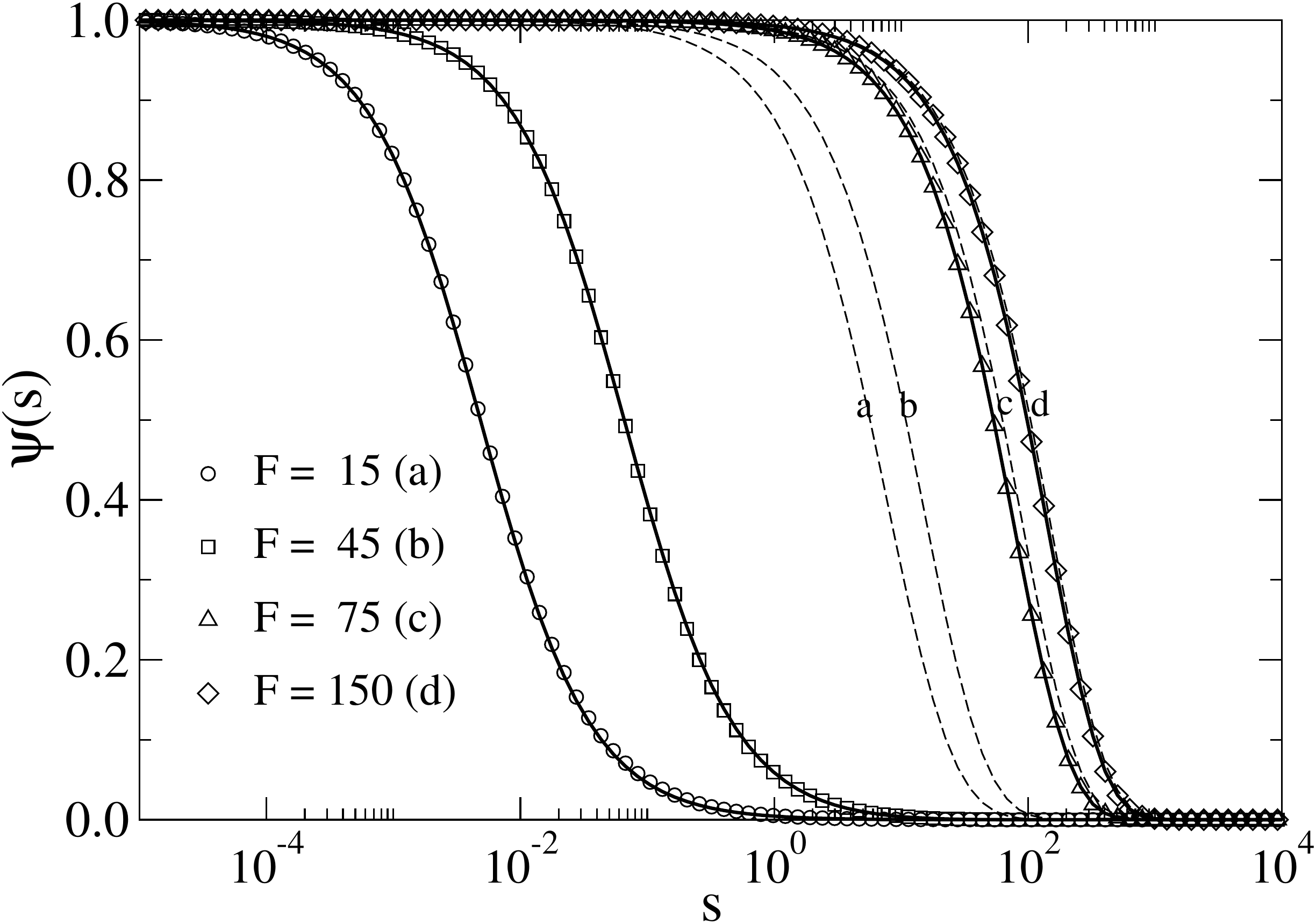}
 \caption{Empirical Laplace transform for electrostatic barrier data,
at $F=15$ (circle), $F=45$ (squares),
$F=75$ (up-triangle), $F=150$ (diamonds).
The solid lines are the Laplace transforms of
$\hat \psi(s)$ for the square barrier.
\textcolor{black}{For each $F$, the equivalent barrier height $\phi$ is estimated by
formula \eqref{eq:phi_equiv}: $\phi(15)\simeq 8.13$, $\phi(45)\simeq 4.21$,
$\phi(75)\simeq 3.14$ and $\phi(150)\simeq 2.365$.}
The dashed curves labelled by a,b,c and d indicate the Laplace
Transform of the Inverse Gaussian computed at fields
$F=15,45,75,150$ respectively. Clearly, the Inverse Gaussian
becomes a reasonable approximations only for high enough force.
\label{fig:distr2}
}
\end{figure}

Further analysis on translocation statistics can be carried out 
by collecting histograms of the first exit time from the $x=1$ boundary.
Since no analytical expressions are available for the $\psi(t)$ 
in the presence of the generic potential, we need to resort to the 
equivalent square-barrier approximation, that
however provides explicit formulas only in Laplace transform space.
Therefore a direct comparison of the normalized histograms with the
approximated results \eqref{eq:phis} requires the Laplace-inversion
$\psi(t) = {\cal L}^{-1}[\psi(s)]_t$, which should be numerically
performed via standard algorithm.
To avoid the iteration of a boring fitting procedure made of a step of
numerical inversion followed by a step of parameter tuning
in Eq.~\eqref{eq:phis}, it is convenient to employ the so called
{\em empirical} Laplace transform~\cite{henze2002goodness} which
for a set of $M$ measured exit times $\{t_k\}_{k=1}^M$ is defined as
\begin{equation}
\psi_e(s) = \frac{1}{M}
\sum_{k=1}^M e^{-s t_k} \, .
\label{eq:empLap}
\end{equation}
In this way, instead of comparing the distributions we compare their
Laplace transform, in other words the comparison between data 
and theory is not done in the time-argument, as natural, but in the 
$s$-argument.  
In Fig.~\ref{fig:distr2}, the symbols represent the $\psi_e(s)$
associated to four sets of exit times at different values of $F$.
The solid lines correspond to Eq.~\eqref{eq:phis} with the barrier
height $\phi$ estimated by inverting Eq.~\eqref{eq:tau3},
\begin{equation}
\phi = \ln\bigg\{ 1 + \frac{F(F\tau -1)}{1 - e^{-F}}\bigg\}~,
\label{eq:phi_equiv}
\end{equation}
where $\tau$ is the numerical value obtained from formula \eqref{eq:tau}.
The agreement between $\psi_e(s)$ and Eq.~\eqref{eq:phis} is striking
and indicates that the shape of the time distribution can be well 
captured by adjusting a step-like barrier, 
regardless of the details of the true potential. 
The dashed lines from the left to the right
refer to the inverse Gaussian computed at fields
$F=15,45,75,150$ respectively. We see that at low and moderate
fields $F=15,45$ the Inverse Gaussian yields a bad representation of
the exit time process.
This comparison between data and theory via the Laplace transform,
suggests that even the simplest correction to the Inverse Gaussian
model is just able to drastically improve the description of the 
translocation time distribution in the regimes of small fields 
where the Inverse Gaussian is known to be not applicable.
To confirm the above scenario, we numerically inverted
Eq. \eqref{eq:phis} 
\textcolor{black}{via the fixed Talbot algorithm \cite{talbot1979accurate}
implemented in 
Mathematica Wolfram 8.0 by Abate and Valk\'o
\cite{abate2004multi} (package 
FixedTalbotNumericalLaplaceInversion.m)}
using the equivalent $\phi$ from Eq.~\eqref{eq:phi_equiv}.

\begin{figure*}[thp]
\includegraphics[clip=true,width=0.93\textwidth,height=6.0cm]
{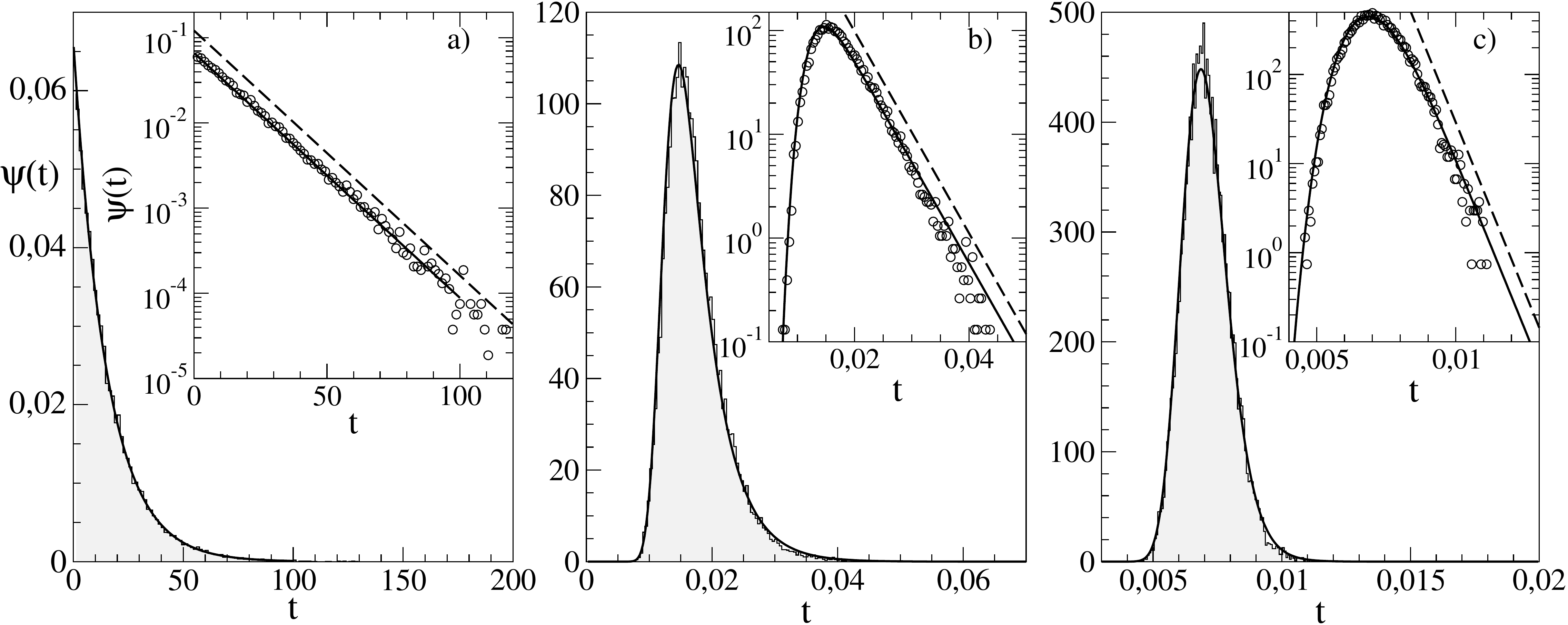}
\caption{Translocation time distributions over the electrostatic barrier,
for three different external loads $F = 15$ a), $F=75$ b) and $F=150$ c).
The shaded areas represent the histograms collected from the arrival time
at $L=1$ of the trajectories generated by the numerical integration
of Eq.\eqref{eq:lang1}. 
\color{black}{The black solid lines are the distribution computed with the 
model of equivalent-square barrier, with $\phi = 8.13$ a), $\phi=3.14$ b) and 
$\phi= 2.365$ c) set from Eq.\eqref{eq:phi_equiv};
consider that no further parameter adjustment is required to fit
perfectly the data.
The insets are the lin-log plots of the
main figures showing the exponential tail of the distributions 
predicted by Eq.\eqref{eq:expcoeff}
(dashed line)}}.   
\label{fig:distr1}
\end{figure*}

The distributions of $t$ for three different forcing are
reported in Fig.~\ref{fig:distr1} where the normalized histograms
are represented by the shaded areas, whereas the solid curve indicates
the theoretical distribution from the equivalent square barrier model.
The first panel refers to the very
low forcing regime ($F\ll F_c$) where the distribution is basically
dominated by the exponential tail.
At low forcing, $F<F_c$, the distribution of panel b) 
is strongly skewed and develops a long tail for large $t$ that yet
differs from the Inverse Gaussian as the waiting time 
before the barrier jumps cannot be neglected.
Finally panel c illustrates a case with $F> F_c$, $\psi(t)$ does not
differ qualitatively from the corresponding Inverse Gaussian with
the same value of $F$, at strong fields indeed, the 
barrier height is drastically reduced and the jump process becomes 
irrelevant.
\textcolor{black}{
The insets of  Fig.~\ref{fig:distr1} report the lin-log plot of the 
main panel data, showing the good agreement between histograms and 
theoretical PdF also in the long-time tails. Moreover, the dashed 
straight lines represent the exponential decay of the tails as 
predicted by the formula 
\eqref{eq:expcoeff} and derived in Appendix \ref{app:LapInv}. 
The perfect alignment of the dashed line with the numerical inversion 
of Eq.\eqref{eq:phis} (solid line) in all the force regimes 
indicates that the behaviour predicted by Eq.\eqref{eq:expcoeff} is an 
exact asymptotic result.} 
The agreement between theoretical predictions and simulation
data is remarkable suggesting the general applicability of the
square barrier model to obtain a first reasonable
correction to the Inverse Gaussian distribution.

\section{Conclusions}
Translocation of molecules through narrow nanopores is often described
as one-dimensional driven diffusion over a energy or free-energy 
profile.
This approach is justified as long as a single reaction coordinate 
is capable of characterizing the transport dynamics while  
the motion along any other direction can be considered to be much faster
\cite{ReactQ}.
Although this assumption may not be true in general, one-dimensional models 
are considered useful mathematical tools to describe the qualitative features 
of the transport phenomenology across narrow pores. 

In this paper, we investigated the diffusion of a single
particle driven by an external constant field along 
a realistic energy profile.
We worked out an analytical expression of the average translocation
time for generic energy profiles, and of the Laplace transform of
the translocation-time distribution over a square barrier.
As a representative case to compare with our theoretical results,
we selected the transport of a positive charged particle through an
$\alpha$HL pore, whose electrostatic potential has been computed 
via the \emph{Poisson} equation.

In order to collect the statistics of the translocation
time at different values of the forcing, we performed numerical 
simulations of the barrier crossing process by solving a Langevin 
equation. 
The average translocation time has been found to be 
in good agreement with the our analytical prediction. 
The energy landscape strongly affects the transport 
at low forcing, determining a non-trivial behavior of the average 
translocation time with the force intensity $F$. 
At high forcing the role of the energy landscape can be neglected, 
thus the translocation process is equivalent to a driven-diffusion 
mechanism without barriers. As a consequence the average translocation 
time shows the quasi-ballistic $F^{-1}$-dependence.     

Also, translocation time distributions clearly reflect the presence of 
these two forcing regimes. At low forcing, where the 
effects of the energy profile are particularly relevant, we were unable 
to derive the exact expression of such distributions, so we 
resorted to the simplest yet still meaningful approximation by     
replacing the actual potential barrier with an appropriate square 
profile. 
In this way, the true Laplace transform of the translocation-time 
distribution over the $\alpha$HL electrostatic barrier is well 
approximated by the corresponding distribution in the model with the 
{\em ``equivalent''} square barrier.
 
Working with the Laplace transform is convenient as it maps a differential
equation problem into an algebraic one. However the main difficulty 
relies in the inversion of the transformation, a step that is often done 
numerically. 
In this respect, we have shown that by employing the {\em empirical Laplace
transform} (\ref{eq:empLap}) of translocation-time raw data, 
the inversion procedure can be skipped, as the comparison between 
theoretical and numerical distributions is equivalent to comparing 
their respective transformations.  
We expect these observations to help the development of
new methods based on empirical Laplace
transforms for the analysis of experimental translocation data.
\textcolor{black}{
Finally, we obtained the analytical decay law of the arrival-time 
distributions at large times for any square-barrier height and any 
explored forcing.
This exponential behaviour $\psi_{\phi}(t) \sim \exp(-\gamma t)$ is 
controlled by the coefficient $\gamma = F^2 e^{\phi}/(1+e^{\phi})^2$
that depends both on the barrier height $\phi$ and the drive $F$.  
}

\appendix \section{Derivation of analytical expression}
\label{app:ana}
In this appendix we derive the formulas \eqref{eq:tau} and \eqref{eq:phis}
from the Smoluchowski equation for a particle in a potential
$G(x) = \bar U(x) - Fx$, where $\bar U(x)$ is significantly non-zero only in
the region $[0,1]$, Fig.~\ref{fig:barr} while $F$ acts over $]-\infty,1]$.
We use natural variables to get a dimensionless equation and we consider
the Laplace transform
\begin{equation}
Y(x,s)=\int_{0}^{\infty}dt P(x,t) e^{-st}\;,
\label{eq:LaplaceP}
\end{equation}
of the probability distribution, which satisfies the equation
\begin{equation}
sY(x,s) - \delta(x) = \frac{\partial}{\partial x}\bigg\{
\frac{\partial Y(x,s)}{\partial x}
+ G^{'}(x) Y(x,s) \bigg\}
\label{eq:LaplaceEq}
\end{equation}
with initial condition $P(x,0) = \delta(x)$.
This problem cannot be solved in a closed form except for some special choice
of $G(x)$, yet, according to \eqref{eqn:meantime}, the 
exact expression of
the average translocation time derives from the solution of the
simplified problem with $s=0$,
\begin{equation}
\tau = \int_{-\infty}^{1} dx Y(x,0)\;.
\label{eq:tau_from_Laplace}
\end{equation}
As it is customary, the presence of the term $\delta(x)$ in \eqref{eq:LaplaceEq} requires
to solve the homogeneous equation
\begin{equation}
\frac{\partial}{\partial x}\bigg\{
\frac{\partial Y(x,0)}{\partial x}
+ G^{'}(x) Y(x,0) \bigg\} = 0
\label{eq:LaplaceEq0}
\end{equation}
in each of the two domains $R_{1}=]-\infty,0[$ and $R_{2}=[0,1]$:
$$
Y(x) =
\begin{cases}
a_0 e^{Fx} & \quad \mbox{if~} x < 0 \\
    e^{-G(x)}[a_1 + b_1 \int_{0}^{x} d\xi e^{G(\xi)} ]
& \quad \mbox{if~} 0 \leq x \leq 1
\end{cases}
$$
where we omit writing $s=0$ for simplicity. The coefficients are determined by
the boundary conditions $Y(-\infty) = Y(1) =0$ and by the two matching conditions
at $x = 0$, resulting from the continuity, $Y(0^+) = Y(0^-)$, and from
integrating both members of Eq.~\eqref{eq:LaplaceEq0} over the interval
$[-\epsilon,\epsilon]$ then taking $\epsilon\to 0$ (see ref.~\cite{risken1996fokker} pp.112-116).
The latter leads to the following current jump
\begin{equation}
-1 = [Y'(x) + G^{'}(x) Y(x)]_{x=0^+} - [Y'(x) + G^{'}(x) Y(x)]_{x=0^-}
\label{eq:currJump}
\end{equation}
which, assuming the continuity of $\bar U^{'}(x)$ and $Y(x)$ in $x=0$, reduces to $-1 = Y'(0^+) - Y'(0^-)$. In
other words, the derivative $\partial_x Y$ presents a discontinuity of magnitude $1$ in $x=0$.
These three conditions determine univocally the coefficients $a_0,a_1,b_1$,
which read:
$$
a_0 = -a_1 = \int_{0}^{1} dq e^{\bar U(q) - F q}\;, \qquad b_1 = -1\;.
$$
Then according to formula \eqref{eq:tau_from_Laplace},
the direct integration of $Y(x)$ over $]-\infty,1]$
yields the result \eqref{eq:tau},
$$
\tau = \frac{a_0}{F} + \int_{0}^1 dx
e^{\bar U(x) - F x} \int_{x}^1 dy e^{-\bar U(y) + F y}\;.
$$

The case with a square barrier in $[0,1]$,
$\bar U(x) = \phi \:\Theta[x(1-x)]$, $\Theta(s)$ being the unitary step function,
has the advantage of being fully tractable even for $s\neq 0$. In this case,
Eq.~\eqref{eq:LaplaceEq} reduces to
\begin{equation}
Y^{''}(x,s)- F Y^{'}(x,s) - s Y(x,s) = 0~,
\label{eq:LaplaceEq2}
\end{equation}
whose solutions in $R_1$ and $R_2$
are linear combinations of $\exp(Fx/2 \pm\sqrt{F^2+4s}/2)$, {\em i.e.}:
\begin{equation}
Y(x,s)=\left\{ \begin{array}{ll}
e^{F x/2} (A_1 e^{q x} + A_2 e^{-q x}) &
x <0 \\
e^{F x/2} (B_1 e^{q x} + B_2 e^{-q x}) & 0\leq x\leq 1
\end{array}\right.
\label{eqn:SolRegion}
\end{equation}
with $q = \sqrt{F^2+4s}/2$.
As before, the four coefficients ${A_1,A_2,B_1,B_2}$ are determined by the boundary conditions:
$Y(-\infty,s) = 0$, $Y(1,s) = 0$ and by the matching at the discontinuity of the potential,
$Y(0^{+},s) e^{\phi} = Y(0^{-},s)$, in which the barrier height $\phi$ appears
\cite{berdichevsky1996one}.
Again, the presence of the $\delta$-function imposes the constraint 
\eqref{eq:currJump}:
$$
-1 = Y'(0^+,s) - F Y(0^+,s) - [Y'(0^-,s) - F Y(0^-,s)] \;.
$$
After simple but tedious algebra, we obtain the following solution
\begin{equation}
Y(x,s)=\left\{ \begin{array}{ll}
 h_1(s)\; e^{F x/2} \exp(qx)  & x < 0 \\
 h_2(s)\; e^{F x/2} \sinh[q(1-x)] & 0 \leq x <\leq 1
\end{array}\right.
\end{equation}
where the coefficients $h_1,h_2$ read
\begin{eqnarray}
h_1(s) = \frac{2 e^{\phi} \sinh(q)} {2q\cosh(q)
+ [2 e^{\phi}q - F(e^{\phi}-1)]\sinh(q)} \nonumber \\
 h_2(s) = \frac{2}{2q\cosh(q) + [2 e^{\phi}q - F(e^{\phi}-1)]\sinh(q)}\;.
\end{eqnarray}
Thus after Laplace transforming Eq.~\eqref{eq:psi1}, which yields $\hat\psi(s) = J(1,s)$,
we have:
\begin{widetext}
\begin{equation}
\hat\psi_{\phi}(s) = \frac{e^{F/2}
\sqrt{F^2 + 4s}} {\sqrt{F^2+4 s} \cosh\left(\frac{1}{2}\sqrt{F^2+4
s}\right) + \left[e^{\phi}\sqrt{F^2 + 4s} - F(e^{\phi}-1)\right]
 \sinh\left(\frac{1}{2}\sqrt{F^2+4s}\right)}
\label{eq:phisapp}
\end{equation}
\end{widetext}
that is the equation \eqref{eq:phis} reported in the section\ref{sec:smo}.
When $\phi = 0$, we simply recover the Laplace Transform of the
Inverse Gaussian \eqref{eq:IGlaplace}
\begin{equation}
\hat\psi_{\phi}(s) = e^{\frac{1}{2} \left(F-\sqrt{F^2+4 s}\right)}\;.
\end{equation}

\section{Laplace inversion
\label{app:LapInv}}
For the sake of completeness we report additional explicit calculations
concerning the inversion of the Laplace transform.
Using the shift property of the Laplace transformation, such that 
$s \to s - F^2/4 $, 
equation \eqref{eq:phisapp} can be recast into the following form
\begin{equation} \psi_{\phi}(t)
= \exp\bigg\{-\frac{F^2t}{4}+\frac{F}{2}\bigg\} 
{\cal L}^{-1}[\hat Q(\sqrt{s})]_t\;,
\label{eq:shift}
\end{equation}
hereafter, ${\cal L}^{-1}[...]_{u}$ the inverse Laplace transform of 
argument $u$. 
Thus, we are left with the simpler function to be inverted
\begin{equation}
\hat Q(\sqrt{s}) = \frac{2\sqrt{s}}{2\sqrt{s}\cosh(\sqrt{s}) +
\left[2e^{\phi}\sqrt{s} - F(e^{\phi}-1)\right]\sinh(\sqrt{s})}\;.
\label{eq:Q_of_sqrts}
\end{equation}
that can be recast in the form
$$
\hat Q(\sqrt{s}) =\frac{2\sqrt{s}}
{e^{\sqrt{s}}[(e^{\phi}+1)\sqrt{s}  -w] - 
 e^{-\sqrt{s}}[(e^{\phi}-1)\sqrt{s} -w]}
$$
where $w = F(e^{\phi}-1)/2$. 
In order to use Laplace transform tables of elementary functions, 
$\hat Q(\sqrt{s})$ can be conveniently re-written as a geometric 
series
\begin{equation}
\hat Q(\sqrt{s}) = 
\frac{2\sqrt{s}\;e^{-\sqrt{s}}}{a\sqrt{s} - w}
\sum_{n=0}^{\infty} e^{-2n \sqrt{s}} 
\left(1 - \frac{2\sqrt{s}}{a\sqrt{s} - w}\right)^n.
\label{eq:serieB}
\end{equation}
with $a=\mbox{e}^{\phi}+1$.

It is immediate to treat the case $F=0 (w=0)$, {\em i.e.} barrier without 
drift, that reduces to invert 
\begin{equation}
\hat Q(\sqrt{s}) = 
\frac{2}{a}  
\sum_{n=0}^{\infty}
e^{-(2n +1)\sqrt{s}}
\left(1 - \frac{2}{a}\right)^n
\end{equation}
leading to the formula
\begin{equation}
\psi_{\phi}^{(0)}(t) = \frac{2}{1+e^{\phi}} \sum_{n=0}^{\infty}
\bigg(\frac{e^{\phi}-1}{e^{\phi}+1}\bigg)^{\!n} \frac{2n+1}{\sqrt{4\pi
t^3}}\;e^{-\tfrac{(2n+1)^2}{4t}} 
\label{eq:nofield}
\end{equation}
When $t$ is large, the sum \eqref{eq:nofield} 
is dominated by the $n=0$ term, thus we 
obtain the asymptotic approximation
$$
\psi^{(0)}_{\phi}(t)
\simeq \frac{2}{1+e^{\phi}} \frac{\exp\{-1/(4t)\}}{\sqrt{4\pi
t^3}}\,.
$$

For the case $F>0$, we perform a further binomial expansion of the term 
raised to the $n$-power in Eq.\eqref{eq:serieB}
\begin{equation}
\hat Q(\sqrt{s}) = 
\sum_{n=0}^{\infty}   
\sum_{k = 0}^{n} 
(-1)^k \binom{n}{k} 
\left(\frac{2\sqrt{s}}{a\sqrt{s} - w}\right)^{\!k+1}
\!\!\!\!e^{-t_n\sqrt{s}},
\label{eq:hatqn}
\end{equation}
for shortness sake, we set $t_n = 2n + 1$ and
$$
\hat Q_{n,k}(\sqrt{s}) = 
\left(\frac{2\sqrt{s}}{a\sqrt{s} - w}\right)^{k+1} e^{-t_n\sqrt{s}}\;.
$$

Applying the Schouten-Van der Pol Theorem (see \cite{Schouten1935} and 
pag.77 of Ref.~\cite{DuffyBook})  
we can invert each term of the sum 
\eqref{eq:hatqn}
with $\sqrt{s}$ replaced by $s$ at the price of solving the integral 
\begin{equation}
{\cal L}^{-1} [\hat Q_{n,k}(\sqrt{s})]_t 
= 
\int_{0}^{\infty} 
\frac{du\,u}{\sqrt{4\pi t^3}} 
e^{-u^2/4t}
{\cal L}^{-1}
[\hat Q_{n,k}(s)]_u \; . 
\label{eq:theorem}
\end{equation}
It can be shown that
\begin{widetext}
\begin{equation}
{\cal L}^{-1}
[\hat Q_{n,k}(s)]_u 
= 
\left(\frac{2}{a}\right)^{k+1}
\left[ 
\delta(u-t_n) + (k+1)\frac{w}{a} 
\Theta(u-t_n)
M\left(k+2,2,\frac{w(u-t_n)}{a}\right)
\right] 
\end{equation}
\end{widetext}
where $\Theta(u-t_n)$ is the unitary step function and 
$M(\alpha,\beta,u)$ indicates the Kummer's confluent Hypergeometric 
function \cite{arfken2005} which for $\alpha = k+2$ and
$\beta = 2$ is known to assume the form $M(k+2,2,x) = e^x P_k(x)$
of a product between an exponential and a polynomial of degree $k$, 
such that $P_0(x) = P_k(0) = 1$.

The above expression plugged into the integral \eqref{eq:theorem} yields
\begin{equation}
{\cal L}^{-1}\!\!
\left[\hat Q_{n,k}(\sqrt{s})\right]_t\! 
=\!\left(\frac{2}{a}\right)^{\!k+1}\! 
\left[
\frac{t_n\,e^{-t_n^2/(4t)}}{\sqrt{4\pi t^3}} + (k\!+\!1)\frac{w}{a} 
I_{n,k}(t) 
\right]
\label{eq:invQ_nk}
\end{equation}
where $I_{p,q}(t)$ represents the integral 
$$
I_{p,q}(t) = \int\limits_{0}^{\infty}\! 
\frac{du(u+t_p)}{\sqrt{4\pi t^3}}
\exp\left\{-\frac{(u+t_{p})^2}{4t} + \frac{w u}{a}\right\} 
P_q\left(\frac{wu}{a}\right).
$$
Combining all the above expressions together and considering that the first 
term of Eq.\eqref{eq:invQ_nk} reconstructs the function \eqref{eq:nofield},
we obtain the final result as a sum  
\begin{equation}
Q(t)
= \psi_{n}^{(0)}(t) + \frac{w}{a} \sum_{(n,k)=0}^{\infty} g(n+k,k) 
I_{n+k,k}(t) 
\label{eq:inversa}
\end{equation}
with coefficients 
$$
g(p,q) = (-1)^q
\binom{p}{q} \left(\frac{2}{a}\right)^{q+1}(q+1)
$$
The substitution of Eq.\eqref{eq:inversa} into \eqref{eq:phisapp} yields 
the first arrival time distribution, which, despite the simplicity of the 
problem, remains quite involved as it amounts to a double series in 
the $k,n$ indexes.
However one can easily derive the simplest nontrivial correction to
Eq.\eqref{eq:nofield} ($F=0$ case) by retaining only the $k=0$
terms in Eq.\eqref{eq:inversa}
\begin{equation}
\psi_{\phi}(t) \simeq 
e^{-F^2 t/4 + F/2}
\left[
\psi_{\phi}^{(0)}(t) + F \frac{e^{\phi}-1}{(e^{\phi}+1)^2}
\sum_{n=0}^{\infty}I_{n,0}(t) 
\right]
\label{eq:low_F}
\end{equation} 

As a final remark, we stress that
asymptotic behaviour of $\psi_{\phi}(t)$ at large times is fully determined,
term by term, by the explicit structure of the integrals $I_{p,q}(t)$.   
The large-$t$ behaviour of $I_{p,q}(t)$ thus can be estimated by evaluating the 
integrand at the saddle point $u^* = 2t w/a  - t_{p}$ of its exponent.
The direct substitution shows that $I_{p,q}(t) \sim \exp[t (w/a)^2]$
which combined with the exponent $-F^2t/4$ from Eq.\eqref{eq:shift},
and recalling that $\psi_{\phi}^{(0)}(t) \sim t^{-3/2}$ is subleading,   
yields the large-time behaviour of the distribution
\begin{equation}
\psi_{\phi}(t) \sim 
\exp\left\{-\frac{e^{\phi} F^2}{(e^{\phi}+1)^2} t\right\}\;.
\label{eq:decay}
\end{equation}
where, we substituted $w = F/2(e^{\phi}-1)$, $a = e^{\phi}+1$.
The analytical result \eqref{eq:decay} is an exact asymptotic
property of the arrival-time distribution of the driven-diffusion
over a square barrier, as it verified with great accuracy in the 
explored range of $F$ by the numerical inversion of Eq.\eqref{eq:phis}, 
in Fig.\ref{fig:distr1}.

\bibliography{refs.bib}
\bibliographystyle{unsrt} 

\end{document}